\documentclass[a4]{article}

\font\scap=cmcsc10 \hfuzz=5cm
\font\scap=cmcsc10

\font\tenmsb=msbm10
\font\sevenmsb=msbm7
\font\fivemsb=msbm5
\newfam\msbfam
\textfont\msbfam=\tenmsb
\scriptfont\msbfam=\sevenmsb
\scriptscriptfont\msbfam=\fivemsb
\def\Bbb#1{{\fam\msbfam\relax#1}}

\newcount\eqnumber
\def\neweq{{(\the\eqnumber)}\global\advance\eqnumber by 1}
\def\eqdef#1{\eqno\xdef#1{\the\eqnumber}\neweq}
\def\newaeq{{(\the\eqnumber a)}\global\advance\eqnumber by 1}
\def\eqdaf#1{\eqno\xdef#1{\the\eqnumber}\newaeq}
\def\eqdisp#1{\xdef#1{\the\eqnumber}\neweq}
\def\eqdasp#1{\xdef#1{\the\eqnumber}\newaeq}

\newcount\refnumber
\def\newref{{\the\refnumber}\global\advance\refnumber by 1}
\def\refdef#1{{\xdef#1{\the\refnumber}}\newref}

\begin{document}

\centerline{\bf Integrable mappings and the notion of anticonfinement}
\par\bigskip
\bigskip{\scap T. Mase} and {\scap R. Willox}\qquad
{\sl Graduate School of Mathematical Sciences, the University of Tokyo, 3-8-1 Komaba, Meguro-ku, 153-8914 Tokyo, Japan }
\par\medskip{\scap A. Ramani} and {\scap B. Grammaticos}
\qquad{\sl IMNC, Universit\'e Paris VII \& XI, CNRS, UMR 8165, B\^at. 440, 91406 Orsay, France}
\par\bigskip
{\scap Abstract}\par
\smallskip
We examine the notion of anticonfinement and the role it has to play in the singularity analysis of discrete systems. A singularity is said to be anticonfined if singular values continue to arise indefinitely for the forward and backward iterations of a mapping, with only a finite number of iterates taking regular values in between. We show through several concrete examples that the behaviour of some anticonfined singularities is strongly related to the integrability properties of the discrete mappings in which they arise, and we explain how to use this information to decide on the integrability or non-integrability of the mapping.

\bigskip
PACS numbers: 02.30.Ik, 05.45.Yv
\par\smallskip
Keywords: mapping, integrability, deautonomisation, singularity, degree growth
\par\bigskip
1. {\scap Introduction}
\par\medskip
When singularity confinement was first introduced as a discrete integrability criterion [\refdef\sincon], the notion of singularity that was considered drew heavily on the corresponding notion for continuous systems [\refdef\physrep]. In the latter case, the existence of a movable critical singularity, i.e. a multivaluedness-inducing singularity the position of which depends on the precise initial conditions, thwarts any attempt to uniquely define a solution. In the case of discrete systems the singularities the confinement criterion refers to arise when at some iteration, determined by the specific initial conditions, one loses a degree of freedom in the solution. In some cases however this degree of freedom is recovered spontaneously through the lifting of a subsequent indeterminacy, a phenomenon which was dubbed singularity confinement. Moreover, it was observed that all mappings integrable through spectral methods do possess the confinement property. On the other hand, generic mappings have unconfined singularities, from which a degree of freedom that was lost is never recovered. These observations are what led us to propose singularity confinement as a heuristic discrete integrability criterion. It turned out to be a very powerful tool in the investigation of integrability yielding a slew of interesting results, prominent among which is the derivation of the discrete analogues of the Painlev\'e equations.

It must be said however that already in the very first work [\sincon] on singularity confinement it was made clear that singularities other than confined or unconfined ones may also exist for a given mapping, for example singularities that are `cyclic' in the sense that they cannot be entered unless they have already been visited in the past. In [\sincon] such singularities were waved aside with the remark that they are not really ``movable'', borrowing the terminology from the continuous case, as one could argue that their position does not really depend on the initial conditions, and these singularities were simply neglected on the tacit assumption that they do not impact on the integrable character of the mapping. We were thus focusing on the confinement property positing that unconfined singularities signal non-integrability, while confined ones are an integrability indicator. 

This last statement, however, requires some clarification. First, unconfined singularities are not necessarily incompatible with integrability (of some sort). Indeed, systems integrable through linearisation may possess unconfined singularities [\refdef\tremblay], in perfect parallel with the continuous case where linearisable systems do not necessarily possess the Painlev\'e property either. Second, there exist systems with confined singularities which are not integrable [\refdef\hiv]. In fact, this might have sounded the death knell for singularity confinement as a discrete integrability criterion were it not for the method of full deautonomisation we introduced in [\refdef\redeem]. We showed there that by considering the confinement constraints on a properly chosen non-autonomous version of a non-integrable mapping with confined singularities, we can obtain a clear indication of its integrability or non-integrability. As a matter of fact, in all cases we examined, we could calculate the exact value of its dynamical degree (or equivalently, of its algebraic entropy).

The groundbreaking work of Halburd [\refdef\halburd] stressed the importance of the `cyclic' singularities mentioned above. These arise in many systems and are of the utmost importance when one wishes to compute the exact degree of the iterates of a mapping that possesses the confinement property, but they do not influence the actual degree {\sl growth} of the mapping, if that growth is exponential. Hence, they have no bearing on the integrable or non-integrable character of the mapping. In fact, in an alternative approach we introduced under the name of ``express method'' [\refdef\rodone], one completely neglects the cyclic singularities without however losing the ability to compute the exact value of the dynamical degree of the mapping, i.e. the ability to decide on its integrability.

While singularities do play a role in the case of higher-order mappings as well, in this paper we prefer to limit ourselves to a well-studied domain where examples abound and where rigorous results do exist, namely that of second order mappings. The paper complements the singularity typology for such mappings,  focusing on yet another type of singularity. There are systems for which singular values persist indefinitely in both the forward and backward iteration with just a finite region of regular values in between. What is interesting in the case of such `anticonfined' singularities, as we shall call them, is that the `orders' of the singular values may exhibit growth (as will  be illustrated in the next section), and that this growth actually gives an indication of the integrability of the mapping. In particular, zero or linear growth in  the orders of the singular values is a possible indication of integrability, while exponential growth signals non-integrability. These statements will be made more precise in the sections that follow.

It should be stressed that although we shall be making pronouncements on the relation between certain types of singularities and the value of the dynamical degree for a given mapping, these observations are very different from those in other studies such as [\halburd] or [\refdef\diller]. For example, in the latter, a classification of the possible degree growths of second order mappings is given in terms of the regularisability or non-regularisability of the mapping, i.e. based on whether or not a mapping can be lifted by a finite number of blow-ups and blow-downs to an automorphism on a certain type of surface. Anticonfined singularities can in fact be of either type, regularisable or non-regularisable. However, as we shall see, this distinction is of little consequence in our approach, as in many cases the growth exhibited by the anticonfined singularity pattern allows us to decide immediately whether the mapping at hand is integrable or not.
\par\bigskip
2. {\scap Enter anticonfinement}
\par\medskip
When the notion of singularity confinement was first proposed,  it was also immediately  recognised that there are many mappings in which singularities simply cannot appear, unless one starts from very special initial conditions, involving  just one degree of freedom, such that when one considers the backward evolution of the mapping the information on this degree of freedom is irretrievably lost, beyond a certain (finite) iteration step. 

Polynomial mappings are perhaps the simplest representatives of this class. Does this mean however that singularity analysis cannot be used to distinguish between a linear (and by definition integrable) mapping such as
$$x_{n+1}+x_{n-1}=\alpha+\beta x_n,\eqdef\aqi$$ 
and, for example, the Henon mapping [\refdef\henon]
$$x_{n+1}+x_{n-1}=\alpha+\beta x_n^2\,,\eqdef\aqii$$
a paradigm of non-integrability?  It turns out  that this  pessimistic  view is unfounded, if one performs the singularity analysis in a way that is slightly more sophisticated than usual. Both equations (\aqi) and (\aqii) can only have a singularity if one starts from the following initial conditions in $\Bbb P^1$: $x_0$  is finite while $x_1$  is taken to be infinitely large. Clearly, given the form of the mappings, the information on the value of $x_0$ is then lost at the first iteration. In practice such initial conditions are implemented by introducing an infinitesimal quantity $\epsilon$, taking $x_1=\epsilon^{-1}$ and keeping only the dominant powers of $\epsilon^{-1}$ in the subsequent iterations.
  
Iterating mappings (\aqi) and (\aqii)  for $x_1=\epsilon^{-1}$ (and a general, transcendental, choice of $\beta$) we obtain the progressions $x_0, \epsilon^{-1}, \epsilon^{-1},\epsilon^{-1},\epsilon^{-1},\dots $ and $x_0, \epsilon^{-1}, \epsilon^{-2},\epsilon^{-4},\epsilon^{-8},\dots\,$, respectively. However, as both mappings describe the same evolution forwards as backwards, it is clear that the initial conditions that give rise to this singularity in the forward direction, also give rise to the same singularity for the backward evolution. These singularities are therefore part of the pattern
$$\cdots, \epsilon^{-1},\epsilon^{-1}, \epsilon^{-1}, \epsilon^{-1}, x_0, \epsilon^{-1}, \epsilon^{-1},\epsilon^{-1},\epsilon^{-1},\cdots\,, $$
for (\aqi) and
$$\cdots, \epsilon^{-8},\epsilon^{-4}, \epsilon^{-2}, \epsilon^{-1}, x_0, \epsilon^{-1}, \epsilon^{-2},\epsilon^{-4},\epsilon^{-8},\cdots\,,$$
for (\aqii).
Moreover, inspection of the exponents of $\epsilon^{-1}$ in the singular parts of the orbit for mapping  (\aqi) shows that the singularity of $x_n$ ($n>1$) always remains of the order of  $\epsilon^{-1}$, whereas for mapping (\aqii) the degree of $\epsilon^{-1}$ grows like $2^{\vert n \vert}$. It is also clear that for each mapping the actual degree growth of $x_n$ (as a rational function of generic initial conditions $x_0$ and $x_1$) should be at least as fast as the growth exhibited by the orders of the respective singularities and hence, that the latter growth actually yields a lower bound for the dynamical degree of these mappings: the values 1 and 2 respectively. In fact, these values turn out to be exactly equal to the dynamical degrees for the respective mappings. Singularities (here `infinities') extending indefinitely {\sl both ways} from a finite set of regular (here finite) values have been encountered before.  The authors of [\refdef\topical] coined the term {\sl anticonfinement} for  such situations.  
  
More examples in the same vein can be easily produced. Let us for instance consider purely multiplicative mappings that can be linearised by taking logarithms. As we explained in [\refdef\viallet], the fact that a transcendental, instead of a birational, transformation is necessary to linearise the mapping makes these systems solvable but not integrable. How is this reflected in the structure of the (in this case) anticonfined singularities of the mappings? Let us start with the reversible rational mapping
$$x_{n+1}x_{n-1}=x_n^k,\eqdef\dqii$$ 
where $k$ is a positive integer. The case $k=1$ will not concern us since its solutions are periodic with period 6, and so we concentrate on the cases $k=2$ and $k>2$. The dynamical degree of the mapping (\dqii) was calculated in [\viallet] where its value was shown to be $(k+\sqrt{k^2-4})/2$. Thus it is equal to 1 for $k=2$, while it is greater than 1 for $k>2$, in agreement with the fact that the $k=2$ case is a linearisable mapping while for $k>2$ chaotic behaviour can be observed numerically. 
The mapping (\dqii) has two singularities, both anticonfined, and in fact dual to one another by swapping the direction of the evolution. If we start, e.g., from $x_{-1}=\epsilon$ and a finite $x_0$, iterating forwards and backwards we obtain the following pattern
$$ {\rm for}\quad k=2\qquad \cdots,\epsilon^4,\epsilon^3,\epsilon^2,\epsilon^1,x_0,\epsilon^{-1},\epsilon^{-2},\epsilon^{-3},\epsilon^{-4},\cdots$$
where the powers of $\epsilon$ and $\epsilon^{-1}$ increase linearly. The second singularity corresponds to the sequence
$$\cdots,\epsilon^{-4},\epsilon^{-3},\epsilon^{-2},\epsilon^{-1},x_0,\epsilon^{1},\epsilon^{2},\epsilon^{3},\epsilon^{4},\cdots$$
exhibiting of course the same linear growth. The case $k=2$ is indeed linearisable: it suffices to introduce $u_n=x_{n+1}/x_n$ in order to reduce (\dqii) to $u_n=u_{n-1}$, the integration of which is trivial.

For $k>2$ on the other hand we find that the anticonfined singularities have rapid growth. For instance,
$$ {\rm for}\quad k=3\qquad \cdots,\epsilon^{21},\epsilon^8,\epsilon^3,\epsilon^1,x_0,\epsilon^{-1},\epsilon^{-3},\epsilon^{-8},\epsilon^{-21},\cdots$$
and the mirrored pattern for the other singularity. For general values of $k$, we find for the powers of $\epsilon$ and $\epsilon^{-1}$ the recursion $d_{n+1}+d_{n-1}=kd_n$ where $d_0=0$ and $d_1=-1$.   The lower bound for the dynamical degree that is obtained from this recursion relation, $(k+\sqrt{k^2-4})/2$, of course coincides with the exact value computed in [\viallet]. 
Note that, contrary to the mappings (\aqi) and (\aqii), the anticonfining patterns for mapping (\dqii)  are not symmetric. 

The situation is even more interesting for the mapping
$$x_{n+1}=x_{n-1}x_n^k,\eqdef\zqi$$ 
where $k$ is again a positive integer. It also has two anticonfined singularities, dual to each other under $\epsilon\to\epsilon^{-1}$. For example, for $ k=1$ we find the pattern
$$\cdots,\epsilon^{8},\epsilon^{-5},\epsilon^{3},\epsilon^{-2},\epsilon,\epsilon^{-1},x_0,\epsilon^{-1},\epsilon^{-2},\epsilon^{-3},\epsilon^{-5},\epsilon^{-8},\cdots,$$
and a second one obtained by $\epsilon\to\epsilon^{-1}$.
One recognizes readily the Fibonacci sequence in the exponents of $\epsilon^{-1}$ for ascending $n$'s, while for the backward evolution the signs of the exponents alternate at each iteration. As this mapping has no other singularities besides the anticonfined ones, one could hope that once again, the golden mean would not simply be a lower bound for the dynamical degree of the mapping but would actually coincide with the latter. This, in fact, turns out to be the case.
\par\bigskip
3. {\scap Anticonfinement and linearisable mappings}
\par\medskip
The occurrence of anticonfined singularities was first noticed in the context of mappings linearisable through birational transformations [\refdef\takenawa]. (The authors of  that reference used the term ``weakly confined'' singularities, but the term ``anticonfined'' proposed in [\topical] is more appropriate). Let us investigate the anticonfined singularities of linearisable mappings on some examples. The mapping 
$$(x_{n+1}+x_n)(x_n+x_{n-1})=a(x_n^2-1),\eqdef\cqi$$
(where $a$ is non-zero and $a\ne1$, lest the mapping become periodic with period 3) is a well-known linearisable system [\refdef\capel]. It belongs to the family known as Gambier mappings [\refdef\gambier] , which can be linearised by rewriting them as a system of two homographic mappings in cascade. Indeed, for mapping (\cqi), if we introduce an auxiliary variable $y_n$ satisfying the homographic equation $a y_n(1-y_{n-1})=1$ and if we relate $y$ to $x$ through $y_n=(x_n+1)/(x_{n+1}+x_n)$, we find (\cqi) upon elimination of $y$.

This mapping has two confined singularities, with patterns $\{\pm1,\mp1\}$, and an anticonfined one with pattern
$$\cdots, \epsilon^{-1}, \epsilon^{-1}, \epsilon^{-1}, x_0, -x_0, \epsilon^{-1}, \epsilon^{-1},\epsilon^{-1},\cdots$$
Note that in this pattern the singularity exhibits no growth at all: it always behaves as $\epsilon^{-1}$, just as in the case of the linear mapping (\aqi). In fact, it is easily checked that the mapping itself exhibits zero growth: starting from generic initial conditions $x_0$ and $x_1$, $x_n$ is always a rational function of degree 2 in $x_0$ and $x_1$ when $n\geq2$. As we mentioned above, the mapping (\cqi) can be written as a set of homographic mappings in cascade, but it turns out that it is actually birationally equivalent to a projective mapping (on ${\Bbb P}^1\times{\Bbb P}^1$), which of course explains the lack of degree growth. To see this  it is convenient to interpret (\cqi) as a mapping on ${\Bbb P}^1\times{\Bbb P}^1$ as
$$(x_{n+1}, y_{n+1}) = \left( y_n, {{a (y^2_n - 1)}\over{x_n + y_n}} - y_n \right),\eqdef\rone$$
after which it is easily seen to be equivalent to the projective mapping 
$$(u_{n+1}, v_{n+1}) = \left( a {u_n - 1 \over u_n}, a{v_n - 1\over v_n} \right),\eqdef\rtwo$$
in the variables $u_n=(x_n+y_n)/(x_n+1)$ and $v_n = (x_n + y_n)/(x_n -1)$. This is not a mere coincidence. In fact, as shown in [\refdef\deserti], all non-periodic second order mappings with bounded degree growth are birationally equivalent to projective mappings on a suitable (Hirzebruch) surface. The delicate point is, of course, finding the proper birational transformation, which is often far more difficult than just obtaining the linearisation in cascade. 

Mapping (\cqi) is special in the sense that it belongs to the QRT [\refdef\qrt] family of mappings. The structure of its anticonfined singularity, however, is not special in the sense that one finds the very same behaviour for Gambier mappings which are not of QRT type. Let us illustrate this with an example. In [\refdef\addit] we derived the mapping
$$(x_{n+1}-x_n)(x_n-x_{n-1})-a(x_{n+1}+x_{n-1}+2x_n)+3a^2=0,\eqdef\yqi$$
and showed that it was of Gambier type, with growth 0, 1, 2, 3, 4, 4, $\cdots$, and that it belonged to the family of mappings proposed initially by Hirota, Kimura and Yahagi [\refdef\hky]. The mapping has one confined singularity $\{a,0,a\}$ and an anticonfined one
$$\cdots, \epsilon^{-1}, \epsilon^{-1}, \epsilon^{-1}, x_0,x_0-a, 2a-x_0,3a-x_0, \epsilon^{-1}, \epsilon^{-1},\epsilon^{-1},\cdots.$$
Again we remark that the singularity exhibits no growth, always behaving as $\epsilon^{-1}$. 
Another family of linearisable mappings, which we call third-kind (linearisable) mappings, does also exist. A very simple example of such a system is 
$$x_{n+1}x_{n-1}=x_n^2-1 ,\eqdef\cqii$$
with confined singularity patterns $\{\pm1,0,\mp1\}$ and an anticonfining pattern
 $$\cdots, \epsilon^{-4},\epsilon^{-3}, \epsilon^{-2}, \epsilon^{-1}, x_0, \epsilon,-1/x_0, \epsilon^{-1}, \epsilon^{-2},\epsilon^{-3},\epsilon^{-4},\cdots ,$$
where the powers of $\epsilon^{-1}$ increase linearly, which implies that the mapping must exhibit a degree growth which is at least linear as well. It is easily checked that the degree growth is indeed exactly linear. This phenomenon is commented upon in [\refdef\mimura] in relation to the results of Diller and Favre [\diller], which tell us that such an anticonfined singularity is necessarily non-regularisable since linear degree growth is incompatible with the singularity structure of a regularisable mapping. Therefore, in the absence of any other singularities, we expect linear growth in the pattern associated with such an anticonfined singularity to imply linear degree growth and thus a dynamical degree equal to 1 for the mapping. Such a mapping, moreover, is linearisable. 
\par\bigskip
4. {\scap A non-integrable mapping with confined singularities}
\par\medskip
In [\refdef\tsuda] the following interesting mapping was studied:
$$x_{n+1}=x_{n-1}\left(x_n-{1\over x_n}\right).\eqdef\eqi$$
This mapping is manifestly not of QRT type [\qrt] and, moreover, it is not integrable, something that can be easily seen by computing the homogeneous degree growth of its iterates. Starting from initial conditions of the form $x_0$ (of degree 0) and $x_1=p/q$ and calculating the degrees in $(p,q)$ of the successive iterates,  we find the sequence 0, 1, 2, 4, 8, 14, 24, 40, 66, 108, 176, 286, 464, 752, 1218, $\dots$. It is straightforward to verify that for $n\ge4$ these degrees obey the recursion relation $d_{n+1}=2d_n-d_{n-2}$. They therefore grow exponentially with a dynamical degree equal to the golden mean $(1+\sqrt 5)/2$.

When studying the singularity structure of (\eqi) one finds that it has two confined singularities, corresponding to the patterns $\{\pm1,0,\infty,\mp1\}$,  and an anticonfined one, which can be entered from initial conditions where $x_1=0$ (for some generic $x_0$): after 4 iterations of the mapping one reaches $(x_4, x_5)=(\infty, \infty)$ which is a fixed point for the mapping (when considered over ${\Bbb P}^1\times{\Bbb P}^1$). However, as opposed to the linearisable cases of section 3, the singularity pattern for this anticonfined singularity exhibits exponential growth,
$$\cdots,\epsilon^{13}, \epsilon^{8},\epsilon^{5},\epsilon^{3},\epsilon^{2},\epsilon,\epsilon, x_0,\epsilon,\epsilon^{-1},-x_0,\epsilon^{-1},\epsilon^{-1},\epsilon^{-2},\epsilon^{-3},\epsilon^{-5},\epsilon^{-8},\epsilon^{-13},\cdots,$$
as the exponents clearly form a Fibonacci sequence. In fact, once we have a succession of infinite values the Fibonacci recursion is a direct consequence of the fact that the dominant part of the mapping is just $x_{n+1}=x_{n-1}x_n$. (And the same argument can be applied mutatis mutandis to the backward evolution where again the zeros obey a Fibonacci recursion).
Since the growth of this anticonfined singularity yields the golden mean as a lower bound for the dynamical degree of the mapping, the mapping is necessarily non-integrable. Moreover, as mentioned above, its dynamical degree actually coincides with the lower bound obtained from the above singularity pattern.

Interestingly, we can reach the same conclusion following the method introduced by Halburd in [\halburd]. In  a nutshell, Halburd's method consists in calculating the degree of the $n$-th iterate of the mapping as the number of preimages of some arbitrary value for $x_n$. Taking the latter to be one of the values that appear in the singularity pattern makes the calculation elementary. In this particular case, it is clear that the value 1 can either appear at some iteration step following regular values or whenever a value $-1$ appeared three steps before. Denoting by $U_n$ the number of spontaneous occurrences of the value 1 in the iteration -- or the number of spontaneaous occurences of the value $-1$, as 1 and $-1$ clearly play the same role --  we find that the degree at iterate $n$, calculated as the number of preimages of the value 1, is given by
$$d_n(1)=U_n+U_{n-3},\eqdef\degi$$ 
where by construction $U_n=0$ for all $n\leq0$ and $U_1=1$.
Similarly, the degree calculated as the number of preimages of 0 is given by
$$d_n(0)=2U_{n-1}+\delta_{n1},\eqdef\degii$$ 
as a zero appears whenever a value 1 or $-1$ appeared one step before and also, just once, in the anticonfined pattern, after a finite initial condition $x_0$. Equating the two expressions (\degi) and (\degii) for the degree, we find that $U_n$ must obey the equation
$$U_n+U_{n-3}=2U_{n-1}+\delta_{n1}.\eqdef\degiii$$ 
The dynamical degree of the mapping is given by the largest root of the characteristic equation for (\degiii), which is precisely the golden mean already obtained by different methods above. In fact the solution of (\degiii) is just $U_n=f_{n+2}-1$ where  $f_n$ obeys the Fibonacci sequence $f_{n+1}=f_n+f_{n-1}$ with intial conditions $f_1=1$ and $f_n=0$ when $n\leq0$. The interesting point here is that, had we tried to compute the degree of the mapping from the number of spontaneous appearances of the value $\infty$, we would have found the obvious contribution $2U_{n-2}$ {\sl plus} a contribution due to the fact that $\infty$ appears an infinite number of times in the anticonfined pattern, with growing exponents. Denoting the contribution of the anticonfined pattern to the number of preimages of $\infty$ at the $n$th iteration by $C_n$, we have
$$2U_{n-1}+\delta_{n1}=2U_{n-2} + C_n,\eqdef\degiv$$
which allows us to calculate this contribution exactly. We immediately find that $C_n=2f_{n-1}+\delta_{n1}$, again given in terms of the Fibonacci recursion. Note that $C_n$ is exactly the sum of two contributions: $f_{n-3}+\delta_{n2}$ and $f_{n}$, corresponding to the exponents of infinity in the anticonfining pattern starting from $x_0$ and $-x_0$ respectively, which clearly shows the link between the growth exhibited by the pattern and the degree growth of the mapping. 
\par\bigskip
5. {\scap Birational transformations and anticonfinement}
\par\medskip
Even when a mapping does not possess anticonfined singularities, there exist cases where such a singularity can be induced by an appropriate birational transformation. Of course, since such a transformation does not modify the degree growth of the birational mapping, one can only ever obtain anticonfined singularities with patterns that exhibit growth compatible with the degree growth of the original mapping.

Up to this point we have only studied autonomous mappings. However, nonautonomous mappings might have anticonfined singularities as well. In this section we study a particular integrable mapping and show the existence of an anticonfined singularity with a growth that  is bounded, even in the case where the deautonomisation leads to a nonintegrable mapping.

We start from the following nonautonomous version of the McMillan [\refdef\mcmillan] mapping
$$y_{n+1} + y_{n-1} = {2 a_n y_n\over y^2_n - 1},\eqdef\mcmi$$
where $a_n \ne 0$, which we shall treat as a mapping on ${\Bbb P}^1\times{\Bbb P}^1$:
$$(x_{n+1}, y_{n+1}) = \left( y_n,\, {2 a_n y_n\over y^2_n - 1} - x_n \right).\eqdef\spmcmil$$
For this mapping a singularity appears when $y_n=\pm1$, which for a generic choice of the function $a_n$ will be unconfined. As is well-known however, in case $a_n$ satisfies $a_{n+1} - 2 a_n + a_{n-1} = 0$, these singularities become confined with singularity patterns $\{ \pm 1, \infty, \mp 1 \}$ and the mapping can be seen to have quadratic degree growth and hence a dynamical degree equal to 1. In fact, it is a special case of a discrete Painlev\'e II equation [\refdef\papafrank].

One can also, for example, choose $a_n$ such that it satisfies the first so-called `late confinement'  condition (cf. [\refdef\lasalope] for a definition) : $a_{n+4} - 2 a_{n+3} + a_{n+2} - 2 a_{n+1} + a_{n} = 0$. In that case the above singularities will still be confined, but the dynamical degree for the mapping (\spmcmil) will be equal to the unique root greater than 1 of the polynomial $\lambda^4 - 2 \lambda^3 + \lambda^2 - 2 \lambda + 1 = 0$, i.e. approximately $1.8832$. The mapping in that case is  non-integrable despite having confined singularities.

For arbitrary $a_n$, the mapping (\spmcmil) has a fixed point $(x_n, y_n)=(0,0)$ with regular Jacobian matrix 
$$	J_n = \left(\matrix{0 & 1 \cr-1 & -2a_n}\right).\eqdef\rthree$$
Defining $z_n=y_n/x_n^2$, an easy calculation then shows that the birationally equivalent mapping
$$(x_{n+1}, z_{n+1}) = \left( x^2_n z_n,\, {2a_n\over x^2_n z_n (x^4_n z^2_n - 1)} - {1\over x^3_n z^2_n}\right),\eqdef\antimac$$
has an anticonfined singularity, without growth, that arises from $(x_n,z_n)=(0,z_0)$ for any choice of the function $a_n$:
$$\cdots,(\epsilon, \epsilon^{-1}),(\epsilon, \epsilon^{-1}),(\epsilon, \epsilon^{-1}),(\epsilon, z_0),
(\epsilon^2, \epsilon^{-3}),(\epsilon, \epsilon^{-1}),(\epsilon, \epsilon^{-1}),(\epsilon, \epsilon^{-1}),\cdots$$
This follows immediately from the linearisation of (\spmcmil) around its fixed point
$$\left( \matrix{x_{n+1} \cr y_{n+1}}\right)\approx J_n\left( \matrix{x_n \cr y_n}\right), \quad
\left( \matrix{x_{n-1} \cr y_{n-1}}\right)\approx J^{-1}_{n-1}\left( \matrix{x_n \cr y_n}\right)
= \left( \matrix{-2a_{n-1} & -1 \cr1 & 0}\right)\left( \matrix{x_n \cr y_n}\right),\eqdef\four$$
from which one has that starting from $x_0 = \epsilon, y_0 = \epsilon^2 z_0$ one obtains
$x_1 = z_0 \epsilon^2 \sim \epsilon^2$,
$y_1 = -\epsilon + {\cal O}(\epsilon^2) \sim \epsilon$ and hence $z_1 \sim \epsilon^{-3}$, after which one has $x_n \sim \epsilon$, $y_n \sim \epsilon$ and therefore $z_n \sim \epsilon^{-1}$ for all $n > 1$. The backward (linear) evolution from $x_0 = \epsilon, y_0 = \epsilon^2 z_0$ yields $x_n \sim \epsilon$, $y_n \sim \epsilon$ and hence $z_n \sim \epsilon^{-1}$ for all $n <0$.

The mapping (\antimac) also has singularities when $x^2 z=\pm 1$, induced by the rational transformation $y_n=z_n x_n^2$ from the singularities of (\spmcmil). For a generic choice of $a_n$ these will be unconfined and in general the mapping (\antimac) will be non-integrable. In the case of the late confinement discussed above for (\spmcmil), the mapping (\antimac) will be nonintegrable despite it having two confined singularities and an anticonfined one with bounded growth, something that can be established in a straightforward way by means of the full-deautonomisation procedure as explained in [\redeem]. However, when $a_n$ satisfies $a_{n+1} - 2 a_n + a_{n-1} = 0$, the above singularities for $x^2 z=\pm 1$ are again confined but the mapping will have quadratic degree growth. Thus we see that while the mapping (\spmcmil) may have exponential degree growth due to the presence of unconfined or (late) confined singularities, the anticonfined pattern always exhibits zero growth. Of course this is not in contradiction with what we are positing in this paper, namely that the growth of the anticonfined pattern only offers a lower bound to the actual degree growth of the mapping.

The example above may appear somewhat contrived since it is constructed from a QRT-type mapping and its deautonomisation. However, as we shall show just below, similar conclusions hold for linearisable mappings as well: whether the other singularities are confined or not does not influence the nature of the anticonfined ones.
The latter can exhibit non-exponential growth even when the mapping has unconfined singularities. 
We can illustrate this by the following example, 
$$x_{n+1} = {(1+a)x_n-(a+x_n^2) x_{n-1}\over1 + a x_n^2 - (1+a) x_n x_{n-1}},\eqdef\wone$$
for arbitrary transcendental $a$, which is a Gambier mapping with degree growth 0, 1, 2, 3, 4, 5, $\cdots$. Its linearisability is obvious when written as a mapping on ${\Bbb P}^1\times{\Bbb P}^1$:
$$(x_{n+1}, y_{n+1}) = \left( { x_n+y_n \over x_n y_n +1}, a y_n \right).\eqdef\cqiii$$
Singularities exist when $y=\pm1$ and both are unconfined. Mapping (\cqiii), as it stands, does not have any anticonfined singularities. However, it is straightforward to construct a birationally equivalent mapping that does have such a singularity. As was the case for the McMillan mapping (\mcmi), based on the observation that (\cqiii) has the fixed point $(0,0)$ with regular Jacobian matrix 
$J = \left( \matrix{1 & 1 \cr 0 &a}\right)$, we introduce the new variable $z_n=x_n/y_n^2$. This yields the mapping
$$( y_{n+1}, z_{n+1}) = \left( a y_n, {z_n+y_n \over a^2 y_n (z_n y_n^3 +1)} \right),\eqdef\cqiv$$
which still has unconfined singularities when $y=\pm1$. However, an anticonfined singularity now arises when $y_n=0$ and $z_n$ is finite but non-zero:
$$\cdots,(\epsilon, \epsilon^{-1}),(\epsilon, \epsilon^{-1}),(\epsilon, z_0),(\epsilon, \epsilon^{-1}),(\epsilon, \epsilon^{-1}),\cdots$$
Again, this follows immediately from the linearisation of (\cqiii) around its fixed point,
$$\left( \matrix{x_{n+1} \cr y_{n+1}}\right)\approx J\left( \matrix{x_n \cr y_n}\right), \quad
\left( \matrix{x_{n-1} \cr y_{n-1}}\right)\approx J^{-1}\left( \matrix{x_n \cr y_n}\right)= \left( \matrix{1 & -1/a \cr0 & 1/a}\right)
\left( \matrix{x_n \cr y_n}\right),\eqdef\wtwo$$
from which one has that starting from $x_0 =  \epsilon^2 z_0$ and $y_0 = \epsilon$ one obtains
$x_n = y_n = {\cal O}(\epsilon)$ for all $n\neq 0$ and hence $z_n \sim \epsilon^{-1}$ except for $n=0$. We remark that here again the anticonfined singularity we deliberately introduced, does not exhibit any growth, although the mapping has unconfined singularities.

There is however one more possibility. Consider the Gambier mapping  
$$x_{n+1}={1\over x_{n-1}}+x_n-{1\over x_n}+a,\eqdef\revi$$
obtained in [\refdef\revisit] by eliminating $y$ from the system $y_n=y_{n-1}+a$, $x_{n+1}=y_n-1/x_n$.
It has a confined singularity pattern $\{0,\infty\}$ and an anticonfined one with bounded growth, namely 
$$\cdots, \epsilon, \epsilon, \epsilon, x_0, \epsilon^{-1}, \epsilon^{-1},\epsilon^{-1},\cdots.$$
Note again the no-growth behaviour of the anticonfined singularity typical of the Gambier mappings.
Still, computing the degree growth of the mapping leads to the sequence 0, 1, 2, 3, 4, 5, $\cdots$. This may appear somewhat astonishing since in section 3 both Gambier mappings with confined singularities had a bounded degree growth. 
However, the difference with those other cases is that the fixed point $(\infty,\infty)$ for (\revi) is not an indeterminate point of the mapping, and is therefore not regularisable. Hence, as shown in [\diller], the degree growth of such a mapping can only be linear or exponential, the former being the case here.
\par\bigskip
6. {\scap Conclusion}
\par\medskip
The aim of this paper was to add another tool to the arsenal of singularity analysis, when used as a discrete integrability detector. Our main tool for the investigation of integrability of discrete systems is, as always, the singularity confinement criterion (which has been promoted to a sufficient one thanks to the introduction of the full-deautonomisation approach [\redeem]). While the singularity confinement approach is most powerful, there exist cases where it cannot be applied, for example when the mapping at hand does not possess `movable' singularities, in the sense explained above. This is for instance the case for polynomial mappings. In such cases our new approach, based on the growth exhibited by the anticonfined singularity patterns, offers a handy criterion for integrability and complements nicely the confinement approach. 

Whereas confinement implies the existence of a singularity for some finite number of iterations, preceded and followed by regular values, anticonfinement is in a sense the mirror image of this situation, where a few regular values (or even a single one) are somehow trapped in the midst of an infinite sequence of singularities. However, a mapping with an anticonfined singularity is not a priori non-integrable. In order to decide on the integrability of the mapping, one should study the propagation of the singularity and assess the growth of its order with each iteration. Based on our results we can summarise the different possible situations as follows.

In the case of an anticonfined singularity with zero growth, two cases have to be distinguished. If all other singularities of the mapping are confined then the system at hand may be integrable or not which must be tested with a different approach, for example by the full-deautonomisation or express [\rodone]  methods. If on the other hand the mapping also has unconfined singularities, then the system can only be integrable if it is linearisable. As is well-known by now, the confinement property is not necessary in the case of linearisable mappings [\tremblay]. 

The case of an anticonfined singularity with non-zero growth is somewhat simpler. If  the growth of the order of the anticonfined singularity is linear then  in the absence of any other singularities we can conclude that we are in the presence of a linearisable mapping. If that is not the case then the linearisability (and therefore integrability)  of the mapping will in general depend on the characteristics of the other singularities. However, if the orders in an anticonfined singularity exhibit a growth faster than linear  (and in fact the only known case is exponential), then the mapping is necessarily non-integrable. Exponential growth exhibited by an anticonfined singularity can therefore be used as a non-integrability indicator.

In many of the cases presented here (and in many more examples we also studied) it turned out that the growth-rate of the anticonfined singularities coincides with the value of the dynamical degree. However, based on the examples of section 5 we do not expect this to be valid in general, and the only thing one can be assured of is that the growth of the anticonfined pattern offers a lower bound to the degree growth of the mapping. The general conditions under which such statements can be rigorously shown are an open problem.
\par\bigskip
{\scap Acknowledgements}
\par\medskip
TM and RW would like to acknowledge support from the Japan Society for the Promotion of Science (JSPS),  through the JSPS grants: KAKENHI grant number 16H06711 and KAKENHI grant number 15K04893. 
\par\bigskip
{\scap References}
\begin{description}
\item{[\sincon]} B. Grammaticos, A. Ramani and V. Papageorgiou, Phys. Rev. Lett. 67 (1991) 1825.
\item{[\physrep]} A. Ramani, B. Grammaticos and T. Bountis, Phys. Rep. 180 (1989) 159.
\item{[\tremblay]} A. Ramani, B. Grammaticos and S. Tremblay, J. Phys. A 33 (2000) 3045.
\item{[\hiv]} J. Hietarinta and C-M. Viallet, Phys. Rev. Lett. 81, (1998) 325.
\item{[\redeem]} A. Ramani, B. Grammaticos, R. Willox, T. Mase and M. Kanki, J. Phys. A 48 (2015) 11FT02.
\item{[\halburd]} R.G. Halburd, Proc. R. Soc. A 473 (2017) 20160831.
\item{[\rodone]} A. Ramani, B. Grammaticos, R. Willox and T. Mase, J. Phys. A 50 (2017) 185203.
\item{[\diller]} J. Diller and C. Favre, Am. J. Math. 123 (2001) 1135.
\item{[\henon]} M. H\'enon, Quart. J. Appl. Math. 27 (1969) 291.
\item{[\topical]} B. Grammaticos, R.G. Halburd, A. Ramani and C.M. Viallet, J. Phys. A 45 (2009) 454002.
\item{[\viallet]} B. Grammaticos, A. Ramani and C.-M. Viallet, Phys. Lett. A 336 (2005) 152.
\item{[\takenawa]} T. Takenawa, M. Eguchi, B. Grammaticos, Y. Ohta, A. Ramani and J. Satsuma, Nonlinearity 16 (2003) 457.
\item{[\capel]} A. Ramani and B. Grammaticos, Physica A 228 (1996) 160.
\item{[\gambier]} B. Grammaticos, A. Ramani and S. Lafortune Physica A 253 (1998) 260.
\item{[\deserti]} J. Blanc and J. D\'eserti, Ann. Sc. Norm. Super. Pisa Cl. Sci. (5) Vol. XIV (2015) 507.
\item{[\qrt]} G.R.W. Quispel, J.A.G. Roberts and C.J. Thompson, Physica D34 (1989) 183.
\item{[\addit]} A. Ramani, B. Grammaticos and R. Willox, J. Math. Phys. 58 (2017) 043502.
\item{[\hky]} R. Hirota, K. Kimura and H. Yahagi, J. Phys. A. 34 (2001) 10377.
\item{[\mimura]} A. Ramani, B. Grammaticos, J. Satsuma and N. Mimura, J. Phys. A 44 (2011) 425201.
\item{[\tsuda]} T. Tsuda, A. Ramani, B. Grammaticos and T. Takenawa, Lett. Math. Phys. 82 (2007) 39.
\item{[\mcmillan]} E.M. McMillan  {\sl A problem in the stability of periodic systems} Topics in Modern Physics. A Tribute to E.U. Condon, ed. E. Britton and H. Odabasi (Boulder, CO: Colorado University Press) (1971)  219.
\item{[\papafrank]} F.W. Nijhoff and V. Papageorgiou, Phys. Lett. 153A (1991) 337.
\item{[\lasalope]} T. Mase, R. Willox, B. Grammaticos and A. Ramani, Proc. R. Soc. A 471 (2015) 20140956.
\item{[\revisit]} B. Grammaticos, A. Ramani and S. Lafortune Physica A 253 (1998) 260.
\end{description}

\end{document}